\documentclass[preprint2]{aastex631}

\newcommand{\rf}[1]{{#1}}

\usepackage{amsmath}
\usepackage{xcolor}

\defcitealias{ibanez-mejia2016}{Paper I}
\defcitealias{ibanez-mejia2017}{Paper II}

\submitjournal{ApJ}

\shorttitle{Gravity vs.\ Magnetic fields.}
\shortauthors{Ib\'a\~nez-Mej\'{\i}a et al.}

\begin{document}

\title{Gravity Versus Magnetic Fields in Forming Molecular Clouds}

\correspondingauthor{Mordecai-Mark Mac Low}
\email{email: jcibanezmejia@gmail.com, mordecai@amnh.org,  \\
 klessen@uni-heidelberg.de}

\author[0000-0002-9868-3561]{Juan
 C. Ib\'a\~nez-Mej\'{\i}a}
\altaffiliation{Current address: ASML, Eindhoven, Netherlands}
\affiliation{I. Physikalisches Institut, Universit\"at zu K\"oln, Z\"ulpicher Str. 77, D-50937 K\"oln, Germany}

\author[0000-0003-0064-4060]{Mordecai-Mark Mac Low}
\affiliation{Dept.\ of Astrophysics, American Museum of Natural
 History, 200 Central Park West, New York, NY 10024, USA}
\affiliation{Center for Computational Astrophysics, Flatiron
 Institute, 162 Fifth Avenue, New York, NY 10010, USA}

\author[0000-0002-0560-3172]{Ralf S. Klessen}
\affiliation{Universit\"at Heidelberg, Zentrum f\"ur Astronomie Heidelberg, Institut f\"ur Theoretische Astrophysik, Albert-Ueberle-Str. 2, 69120 Heidelberg, Germany}
 \affiliation{Universit\"at Heidelberg, Interdisziplin\"are Zentrum
 f\"ur Wissenschaftliches Rechnen, Im Neuenheimer Feld 205, 69120 Heidelberg, Germany}

\begin{abstract}
  Magnetic fields are dynamically important in the diffuse interstellar medium. Understanding how gravitationally bound, star-forming clouds form requires modeling of the fields in a self-consistent, supernova-driven, turbulent, magnetized, stratified disk. We employ the FLASH magnetohydrodynamics code to follow the formation and early evolution of clouds with final masses of 3--8$\times 10^3 M_{\odot}$ within such a simulation. We use the code's adaptive mesh refinement capabilities to concentrate numerical resolution in zoom-in regions covering single clouds, allowing us to investigate the detailed dynamics and field structure of individual self-gravitating clouds in a consistent background medium. Our goal is to test the hypothesis that dense clouds are dynamically evolving objects far from magnetohydrostatic equilibrium. We find that the cloud envelopes are magnetically supported with field lines parallel to density gradients and flow velocity, as indicated by the histogram of relative orientations and other statistical measures. In contrast, the dense cores of the clouds are gravitationally dominated, with gravitational energy exceeding internal, kinetic, or magnetic energy and accelerations due to gravity exceeding those due to magnetic or thermal pressure gradients.  In these regions field directions vary strongly, with a slight preference towards being perpendicular to density gradients, as shown by three-dimensional histograms of relative orientation.

\end{abstract}

\section{Introduction}
\label{sec:introduction}

Understanding the relative importance of magnetic fields and self-gravity in the formation, evolution, and collapse of molecular clouds, as well as their envelopes, is critical for understanding star formation \cite[see, e.g.][]{mac-low2004, mckee2007, krumholz2019, chevance2020}. The magnetic field in the diffuse interstellar medium (ISM) is deduced to have a value of 3--5~$\mu$G from observations of H~{\sc i} Zeeman splitting \citep[e.g.][]{heiles2004}, which gives the line-of-sight field strength, and starlight polarization \citep[e.g.][]{mathewson1970,heiles1976}, which traces the projected orientation of the field in the plane of the sky \citep{lazarian2007, andersson2015}. The ionization of the diffuse gas is high enough even in neutral regions for the magnetic field to be effectively coupled to the gas \cite[][]{tielens2010, draine2011, klessen2016, girichidis2020}. As a result, uniform gravitational collapse will result in rapid growth of the field, preventing further collapse long before stars form \citep{mestel1956,Mouschovias1976NoteClouds}. This led to the proposal that collapse and star formation could only occur through the mediation of ambipolar diffusion \citep[ion-neutral drift;][]{mestel1956, mouschovias1977, Shu1977Self-similarFormation}.

Molecular Zeeman splitting observations reveal magnetic field strengths for density ranges depending on the observed tracer molecule \citep[e.g.][]{crutcher1975,falgarone2008}. The inferred values are typically insufficient to support the clouds against collapse, contradicting the expectation from ambipolar diffusion \citep{crutcher1999,crutcher2012}. \citet{padoan1999} showed that the observational characteristics of dense clouds were consistent with super-Alfv\'enic turbulence within them, leading to the gravoturbulent model of magnetized turbulence balancing gravitational collapse \citep{mac-low2004}. More recent observations have emphasized the filamentary nature of the clouds \citep[e.g.][]{andre2014} and have investigated the role of magnetic fields in filament formation \citep[e.g.][]{planckxxxv2016,monsch2018,fissel2019,soler2019a}, raising the question of whether the fields align with the filaments, as might be expected for weak fields, or are perpendicular, as expected for strong fields. 

The observed turbulent motions cannot be driven from within the clouds \citep{brunt2009} but are instead likely associated with the very process of cloud assembly from the tenuous ISM \cite[][]{Klessen2010Accretion-drivenDisks}. \citet[][hereafter \citetalias{ibanez-mejia2016}]{ibanez-mejia2016} demonstrated that the supernova-driven turbulence in the diffuse ISM provides insufficient momentum to the dense clouds to drive the observed level of turbulence within them. Instead, they found that the observed motions can be driven by gravitational collapse, as suggested by \citet{ballesteros-paredes2011}, leading to the conclusion originally proposed by \citet{zuckerman1974}. This leads to the scenario of molecular cloud formation during global hierarchical collapse \citep[e.g.][]{elmegreen2000,vazquez-semadeni2019}, promptly followed by destruction of the clouds by stellar feedback \citep[e.g.][]{dale2015,haid2019,wall2020,grudic2021}.  \rf{\citet{mac-low2017} demonstrated that Toomre instability of the gas disk \citep{goldreich1965} can form dense gas sufficiently quickly to replace clouds promptly destroyed by feedback in order to maintain the observed dense gas fraction in a steady state.}

What magnetic field structure does global hierarchical collapse produce? Is this consistent with observed field properties? In this paper we study these questions using high-resolution, zoom-in simulations of individual clouds drawn from the kiloparsec-scale simulations of \citetalias{ibanez-mejia2016}. The dynamics of accretion and collapse in these objects was described in \citet[][hereafter \citetalias{ibanez-mejia2017}]{ibanez-mejia2017}, while their velocity structure functions were analyzed by \citet{chira2019}. Similar zoom-in simulations were described by \citet{seifried2017} and \citet{girichidis2021}, while their field structure was investigated by \citet{seifried2020} and \citet{girichidis2021}.

We describe the simulations in Sect.~\ref{sec:method}. We determine the physical state of the clouds in Sect.~\ref{sec:physics} using measurements of the sonic and Alfv\'enic Mach number, the different components of the energy, and the forces acting on each parcel of gas, all as a function of density. We then compare our models to observations in Sect.~\ref{sec:observ}, focusing on the relationship between line-of-sight magnetic field and density, and the alignment between magnetic field directions and density gradients as a function of density. 
 
The analysis scripts \rf{and intermediate data products} leading to the results presented here have been deposited in the digital repository of the American Museum of Natural History \citep{doi-cite}.  The underlying simulation results required to run the scripts are available from the same repository \citep{data-cite}, and are also available through the Catalogue of Astrophysical Turbulence Simulations \citep{burkhart2020}.

\section{Numerical simulations}
\label{sec:method}
We analyze models of gravitationally collapsing clouds embedded in a kiloparsec-scale model of three-dimensional, stratified, magnetized, supernova-driven turbulence simulated in the FLASH v4.2.2 adaptive mesh refinement code \citep{fryxell2000,dubey2012}. The computational domain of the embedding simulation is $1\times1\times40$~kpc$^3$, with a background gravitational potential representing a stellar disk and dark matter halo centered vertically. Supernovae (SNe) are set off at the Galactic rate \citep{tammann1994} with distributions characteristic of Type Ia SNe, and both isolated and clustered core-collapse SNe. Diffuse far-ultraviolet photoelectric heating and radiative cooling are both included, as is an ideal treatment of the magnetic fields. The details of this simulation, including the initial conditions, are described in \citetalias{ibanez-mejia2016}. 
 
After establishing an equilibrium background flow at 0.95~pc resolution in the disk midplane ($|z| < 300$~pc), self-gravity was implemented using the FLASH multigrid solver \citep{daley2012}. At this point we have a well-established, multiphase ISM, vertical gas stratification, and a galactic fountain up to $\pm 20$~kpc \citep{Hill2012VerticalMedium,Walch2015TheISM,Girichidis2016TheOutflows}. So far, 7,515 SNe have exploded in the simulation, and they continue to be injected in the subsequent evolution presented in this work. Our analysis is performed from the moment $t_{\rm SG} = 0$ when self-gravity is turned on.

Three clouds were chosen for simulation at much higher resolution. The simulation was restarted prior to the formation of those clouds, with a refinement rule to increase resolution in any region where the Jeans length was resolved with less than four cells, down to a resolution of 0.06 or 0.1~pc. For the latter, this corresponds to resolving a density of $8 \times 10^3$~cm$^{-3}$ for gas at 10~K (Eq.~(15) of \citetalias{ibanez-mejia2017}). Our chosen clouds had masses in the simulation without self-gravity of $3 \times 10^3$, $4 \times 10^3$, and $8 \times 10^3$~M$_{\odot}$; we identify them as M3, M4, and M8 hereafter. The first two were resolved to 0.06~pc, whereas M8 was only resolved to 0.1~pc. The clouds continue to accrete mass throughout their lifetimes. Further details of the high-resolution, zoom-in, re-simulation, and evolution of the hierarchically collapsing clouds are found in  \citetalias{ibanez-mejia2017}.

\begin{figure*}[!h]
\centering 
 \includegraphics[width=1.0\textwidth]{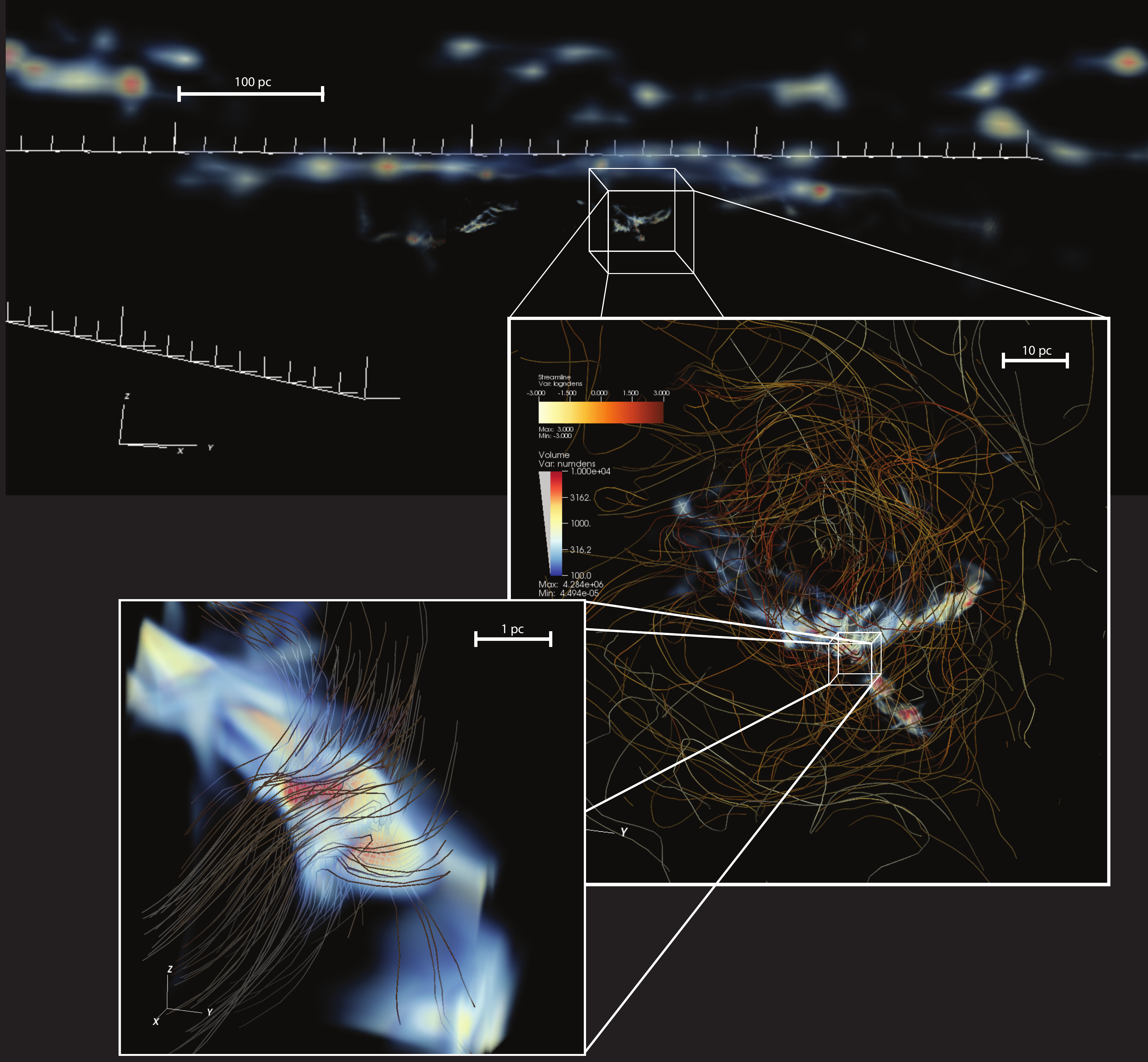} 
\caption{Three-dimensional rendering at $t=5$~Myr of {\em{(Top to
 bottom)}} the galactic midplane for altitudes up to $\pm50$~pc
 above and below the midplane \citepalias{ibanez-mejia2016}, close-up
 view of cloud M3 \citepalias{ibanez-mejia2017}, including magnetic
 field lines for a volume of $50$~pc$^{3}$, and a closer view of a collapsing core with the magnetic field lines crossing a plane with area $2$~pc$^{2}$ centered in the core center of mass. \label{fig:3D_view} } 
\end{figure*} 

As illustration, Figure \ref{fig:3D_view} shows a three-dimensional rendering of all the gas above $100$~cm$^{-3}$ near the midplane, with the location of the high-resolution cloud M3, a close-up of the cloud region, and a second close-up of a dense, collapsing core within the cloud. The zoom-in view includes 125 magnetic field lines that are highly tangled and twisted, penetrating the density structure down to the massive, collapsing cores.

We stop the simulations at an evolutionary time of $t_{\rm SG}\approx10$~Myr, as it is expected that by this time massive stars must have already formed and be feeding back energy in the form of radiation, winds, and SN explosions, which should influence not only the cloud properties but its environment. As we do not include self-consistent star formation and feedback in our simulations, running for longer would lead to unphysical results.

\section{Physical Analysis}
\label{sec:physics}

In order to determine which physical processes dominate the dynamics of the gas in clouds and their envelopes during their formation, evolution, and collapse, we divide our analysis into three parts. First, in Sect.\ \ref{subsec:velocities} we consider the mean velocities at different densities to allow us to determine where
transitions in the sonic and Alfv\'enic Mach numbers of the flow occur. Second, in Sect.~\ref{subsec:Energetics} we compare the distribution of the energy density of the gas in its different forms---magnetic, kinetic, thermal, and gravitational---as a function of the gas density, in and around the collapsing clouds. Third, in Sect.~\ref{subsec:accelerations} we compute the actual accelerations acting on cloud gas to directly determine which forces dominate the flow in each density regime, confirming the results suggested by the earlier analyses. 

The cloud is defined by a number density contour of $n_{\rm c} = 100$~cm$^{-3}$ and a gravitational binding criterion for nearby fragments \citepalias{ibanez-mejia2017}.
  \rf{A 100~pc cube containing the cloud}
is followed through the larger simulation domain
   \rf{and projected onto a uniform Cartesian grid for analysis}
using the {\tt yt} toolkit \cite{turk2011}, as described in \citetalias{ibanez-mejia2017}.

\subsection{Velocities}
\label{subsec:velocities}

We compute three velocities as a function of number density $n$ in the  frame of the cloud \rf{on the uniform grid constructed from the original adaptive grid.} These are the  turbulent rms velocity  $v_{\rm rms}$, the sound speed  $c_{\rm s}$, and the Alfv\'en velocity  $v_{\rm A}$. For each cell $i$ we \rf{compute} 
\begin{align}
v^2_{{\rm \rf{rel}}, i} & = \left(\mathbf{v}_i - \mathbf{v}_{\rm c}\right)^2 ,\\ 
c^2_{{\rm s}, i} & = \gamma k_{\rm B} T_i / \mu  \;, \\
v^2_{{\rm A}, i} &= B_i^2  / (\rf{4} \pi \rf{\rho}) \;.
\end{align}
Here,  $\mathbf{v}_i$, $\mathbf{B}_i$, \rf{$\rho$,} and $T_i$ are the local velocity, magnetic field, \rf{density,} and temperature in the cell, respectively,  and $\mathbf{v_{\rm c}}$ is the center of mass  velocity of the cloud under consideration.  Furthermore, $k_{\rm B}$ is the Boltzmann constant, and we take  $\gamma = 5/3$ (consistent with either monatomic gas or molecular gas  too cold to excite rotational and vibrational levels). We also adopt a mean mass per particle $\mu = 1.3 \,m_{\rm p}$, with $m_{\rm p}$ being the  proton mass. Finally, we subdivide the number density range covered by the simulation into 100 logarithmic bins of size $\delta (\log n)$ between $n = 10^{-2}$~cm$^{-3}$ and  $10^{4.3}$~cm$^{-3}$.  

We obtain the average values as a function of number density $n$ by summing over all $N$ cells $j$ in the range from  $\log n-\delta (\log n)/2$ to  $\log n+\delta (\log n)/2$,
\begin{align}
v_{{\rm rms}}(n) & = \left( \sum_j v^2_{{\rm \rf{rel}}, j} / N \right)^{1/2} ,\\ 
c_{{\rm s}}(n) & = \left( \sum_j c^2_{{\rm s}, j} / N\right)^{1/2}  , \\
v_{{\rm A}}(n) &= \left( \sum_j v^2_{{\rm A}, j} / N\right)^{1/2} .
\end{align}
We also compute the 25th and 75th percentile values to indicate the amount of variation at each number density.

\rf{In this analysis we have neglected to remove any uniform rotation of the cloud \citep[as was done, for example, by][]{federrath2016}, although we have subtracted the bulk motion.  Because the analyzed cube extends well beyond the cloud, subtracting off an average rotation across the cube is unlikely to capture the dynamics of the cloud itself.  In order to assess how much of the rms velocity can be attributed to uniform rotation, we further compute the mass-weighted average rotation of all gas with $n > 100$~cm${-3}$ in the analysis cube.}

\rf{To do this, we start by computing, for each zone $q$ with mass $m_q = \rho_q \delta V_q$ lying within the high-density region, the components $(L_{x,q}, L_{y,q}, L_{z,q})$ of the angular momentum vector
  \begin{equation}
    L_{i,q} = m_q \sum_{j,k} \epsilon_{ijk} v_{j,q} r_{k,q} ,
  \end{equation}
  where $\epsilon_{ijk}$ is the Levi-Civita symbol.  The net angular momentum of the cloud is then ${\mathbf{L}} = \sum_q {\mathbf L_q}$.  We similarly compute the mass $M_{\rm cl} = \sum_q m_q$ and the spherical radius
  \begin{equation} R_{\rm cl} = \left[ (3/4 \pi) \sum_q \delta V_q\right]^{1/3}. \end{equation}
The average rotation velocity is then $v_{\rm rot} = |{\mathbf L}| / (M_{\rm cl} R_{\rm cl})$.
}

The variation of these velocities as a function of density and time shows how the system reacts as the dense gas contracts. The two  dimensionless values that determine its behavior are the sonic Mach  number ${\cal M}= v_{\rm rms}/c_{\rm s}$ and the Alfv\'enic Mach number ${\cal M_{\rm A}} = v_{\rm rms} / v_{\rm A}$.  Supersonic turbulence with ${\cal M}>1$, characteristic of the ISM  \citep{mac-low2004}, leads to large density  fluctuations, as isothermal shocks produce density increases  $\Delta n \propto {\cal M}^2$. Supersonic but sub-Alfv\'enic flows can compress gas along field lines, but cannot strongly influence the  field \cite[see e.g.,][]{beattie2021}. Flows that are both supersonic and super-Alfv\'enic can also compress gas perpendicular to magnetic fields, leading to strong  fluctuations of the magnetic field strength, as the field gets  compressed by the shock along with the gas \cite[see, e.g.,][]{federrath2008, federrath2010b}.
  
\begin{figure*}[t]
\centering 
\includegraphics{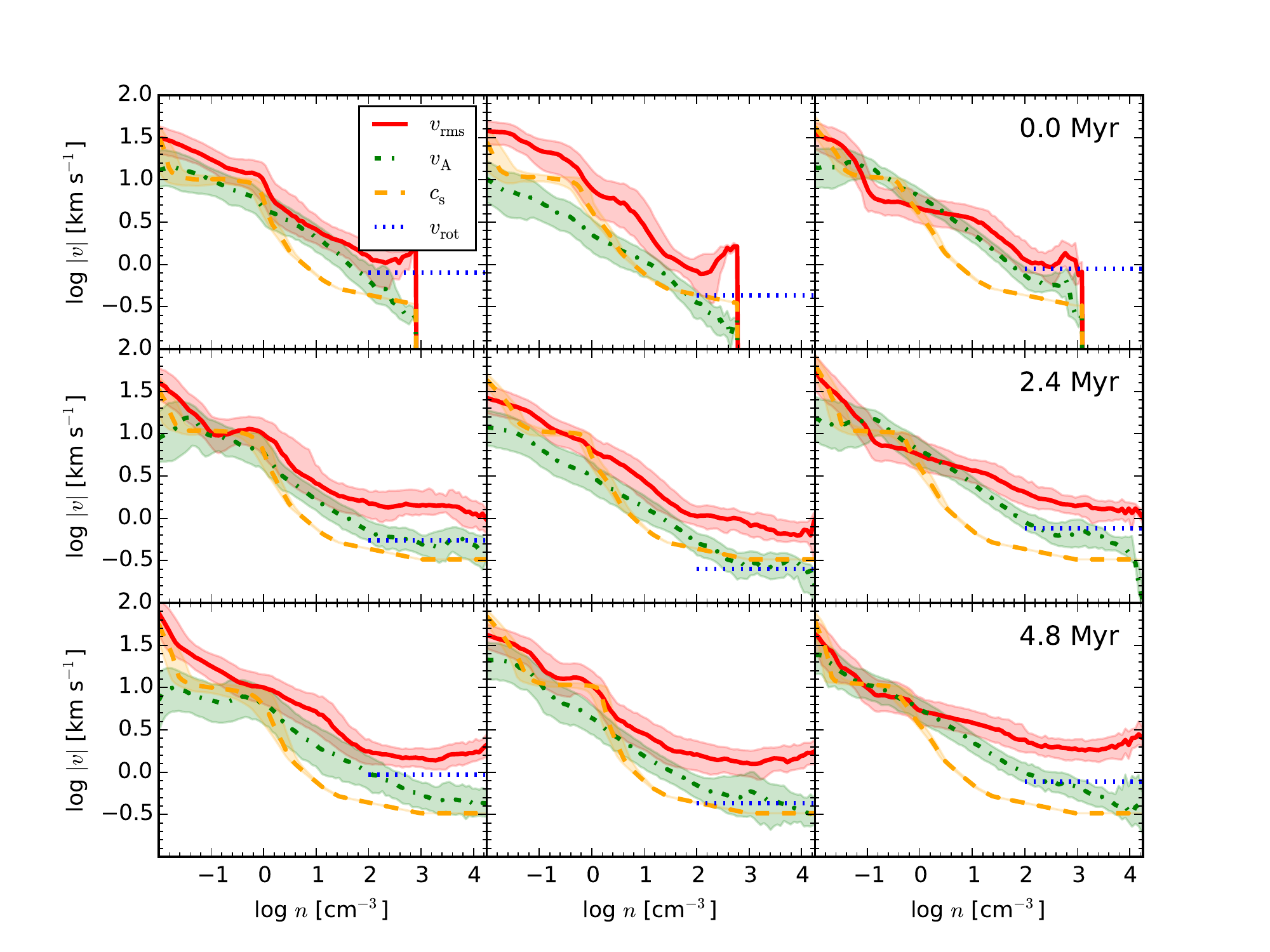}
\caption{Mean rms velocity $v_{\rm rms}$ ({\em red solid}), sound
 speed $c_{\rm s}$ ({\em orange
 dashed}), and Alfv\'en velocity $v_{\rm A}$ ({\em green dash-dotted}) for clouds
 M3 ({\em left}), M4 ({\em center}), and M8 ({\em right})
 as a function of gas density within a ($100$~pc)$^{3}$ box, centered in the cloud's center of mass.  \rf{The average rotational velocity of the region with $n > 100$~cm$^{-3}$ is also shown ({\em blue dotted}).}
The panels correspond to evolutionary times of 0, 2.5, and 4.\rf{8}~Myr (from top to bottom)
since the moment self-gravity was included. The lines show the
volume-weighted medians, while the shaded areas show the
central 50\% of values in each density bin.
\label{fig:characteristic_speeds} }
\end{figure*}

Figure \ref{fig:characteristic_speeds} shows the characteristic velocities of the system as a function of the density, at three evolutionary times of the cloud. At each time, the rms velocity exceeds the sound speed of the system at all but the very lowest densities in the hot, diffuse medium. The flow ranges from mildly supersonic (${\cal   M}\approx1$) in the diffuse ISM at number densities $10^{-2} < n <1$~cm$^{-3}$, to hypersonic, ${\cal M}\approx 5$--10 for  $n > 10$~cm$^{-3}$.  This reproduces the expected general behavior of the turbulent ISM with supersonic shocks permeating the gas at all scales and densities \citep{mac-low2004, padoan2011, vazquez-semadeni2015, klessen2016, girichidis2020}. \rf{As density increases, the transition to hypersonic Mach numbers has been shown to occur at roughly the same density as the transition from atomic to molecular gas \citep{mandal2020}.}


The development of the Alfv\'en number, on the other hand, shows the changing nature of the cloud as self-gravity takes hold. At all times the rms velocity remains trans-Alfv\'enic with ${\cal M_{\rm A}}\approx 1$ at $n<10^{2}$~cm$^{-3}$. At $t = 0$, the rms velocity at the highest densities remains flat, while the Alfv\'en velocity drops, suggesting constant field strength as density increases, consistent with compression primarily along field lines. However, as the density goes up and gravitational acceleration becomes stronger,  the Alfv\'en velocity increases again.

The flow velocity grows more quickly, suggesting \rf{gravitational} collapse \rf{has become the dominant process, producing flows} strong enough to compress the field lines kinematically and increase the magnetic flux in the core region. \rf{Comparing the flow velocity to the rotational velocity shows that although rotation is present, it is not the dominant component of the rms velocity at the highest densities.} The \rf{collapse results in} strongly super-Alfv\'enic flows with ${\cal M_{\rm A}} \gg 1$ at $n>10^{2}$~cm$^{-3}$. Such super-Alfv\'enic flows in the dense regions of clouds were already proposed by \citet{padoan1999} based on the morphology and dynamics of observed and simulated clouds.
 
\subsection{Energetics}
\label{subsec:Energetics}

To more directly examine the relationship between thermal and magnetic pressure and gravitational potential, we measure the different forms of energy as a function of number density over time. The kinetic, thermal, and magnetic energy  density in each cell $i$ is
\begin{align}
e_{{\rm k}, i} & = (1/2) \mu n_i (v_i-v_{\rm c})^2 , \label{e:k}\\ 
e_{{\rm th}, i} & = n_i k_{\rm B} T_i , \label{e:th}\\
  e_{{\rm m},i} &= B_i^2 / (8 \pi
                  )\;. \label{e:m}
\end{align}
To calculate the gravitational energy density of the cloud, we first must determine the background potential $\Phi_0$ within which it is embedded. We calculate this as the mean cell-weighted value of the potential in the cells with $n_i < 10$~cm$^{-3}$ in a 100~pc box centered on the cloud. The gravitational energy in cell $i$ is then 
\begin{equation} 
e_{{\rm pot},i} = \mu n_i (\Phi_i - \Phi_0) . \label{e:g}
\end{equation}
We  find the distribution of the kinetic, thermal, magnetic, and gravitational energy densities
in each number density bin $\log n_k$, similar to the velocities in Sect.~\ref{subsec:velocities}, by calculating the values in cells $j$ in the range from $\log n_k-\delta (\log n_k)/2$ to  $\log n_k+\delta (\log n_k)/2$. 
\begin{figure*}[t]
  \centering
   \includegraphics{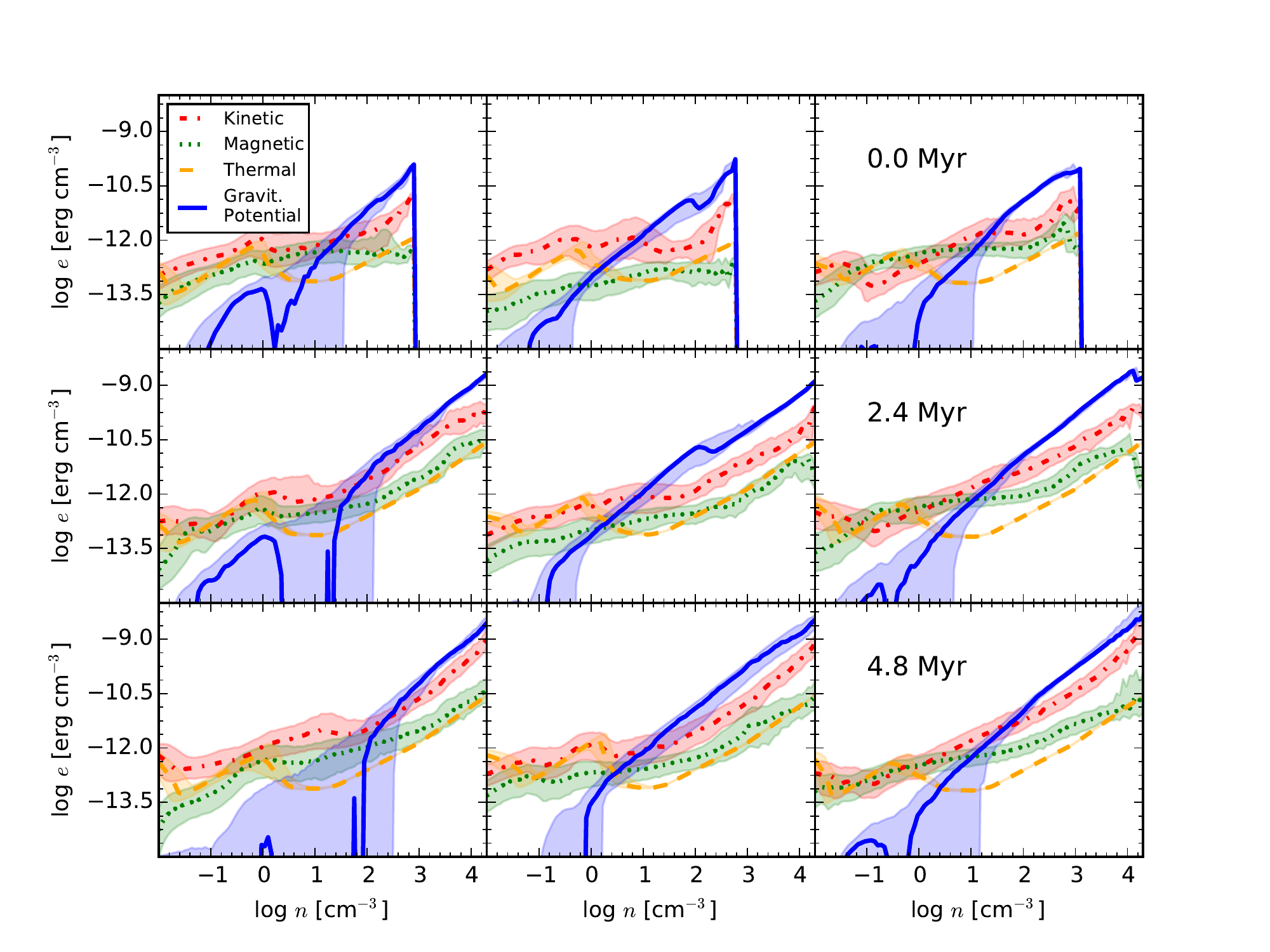}
\caption{
 Energy density in gravitational potential ({\em blue}), magnetic ({\em
 green}), thermal ({\em dashed orange}), and kinetic ({\em red})
 forms as a function of density in the gas within a ($100$~pc)$^{3}$
 box, centered on the cloud's center of mass for clouds M3 ({\em left}), M4 ({\em center}), and M8 ({\em right}) for three snapshots at $t=0$, 2.4, and
 4.8~Myr since the time self-gravity has been turned
 on.  The lines show the
volume-weighted medians, while the shaded areas show the
central 50\% of values in each density bin.\label{fig:Energetics}} 
\end{figure*}
Figure~\ref{fig:Energetics} shows the evolution in time of the median value of each energy density along with the 25th and 75th percentage values as a function of density for each of the three clouds. This complements the picture offered by the mean velocities of the physical processes affecting the cloud dynamics.

   The thermal energy peaks at a density of $n\approx0.1$~cm$^{-3}$ and decreases at higher densities as temperatures drop. The magnetic energy is lower than the kinetic energy at all times, and is maintained at a fraction of $0.1-0.3$ of the total kinetic energy, probably due to the saturation of the small-scale, turbulent dynamo in our simulation box \citep{Balsara2004AmplificationTurbulence,Meinecke2014TurbulentWaves,gent2021}.  The small-scale dynamo drives flux growth at wavelengths close to the dissipation scale, which is  determined in these models by the numerical resistivity.

As the simulation evolves, the gas with $n \gtrsim 100$~cm$^{-3}$ is dominated by gravitational potential energy, implying that the clouds are gravitationally bound and in a state of hierarchical contraction, not supported by turbulent, magnetic, or thermal energy. As the clouds contract, gas flows towards local centers of collapse. These increase in density, resulting in a rise of the gravitational potential energy density at higher densities, followed by a rise of the kinetic energy density as the gas gains velocity falling down the gravitational potential wells. Still, the kinetic energy density within dense regions remains at a fraction 0.2--0.5 of the gravitational potential energy density, closely following its mass density distribution. This is expected for gravitational collapse and supports our hypothesis \citepalias{ibanez-mejia2016} that hierarchical contraction drives the fast, non-thermal, motions of the dense gas in the cloud. The magnetic field energy also increases as the cloud contracts, as field lines are compressed in local centers of collapse. However, magnetic energy density does not grow as fast as kinetic energy density, suggesting that a significant amount of gas moves along field lines.

At densities corresponding to the outer envelope of the cloud, i.e.\ $n = 10$--100~cm$^{-3}$, the kinetic energy contributes the majority of the total energy followed by the magnetic energy. This suggests that the accretion onto the cloud proceeds generally along the magnetic field lines without compressing them strongly.

At every time in the evolution of the simulation, the kinetic energy dominates the energy of the gas below $n \approx 10$~cm$^{-3}$,
as expected in the mildly supersonic and super-Alfv\'enic diffuse ISM.  Large variations in the kinetic, thermal, and magnetic energies are observed. In most cases, the warm medium with $0.1 < n < 10$~cm$^{-3}$ has a higher magnetic energy than thermal energy as expected \citep[e.\ g.][]{heiles2005}, while in the lower density hot medium thermal energy tends to dominate.  
Equipartition is sometimes reached, but is by no means universal, with order of magnitude deviations also occurring in this diffuse gas.
This occurs because we perform our analysis in a $(100$~pc$)^{3}$ box, with open boundaries to a turbulent, multiphase ISM. The inflow and outflow of blast waves and rarefied gas from nearby SN explosions is the main cause for these large fluctuations.

\subsection{Accelerations}
\label{subsec:accelerations}

Finally, in order to confirm our results from the previous subsections, we determine the acceleration $\mathbf{a}$ of the gas in each cell $i$ from the pressure and gravitational gradients and from the magnetic force, 
\begin{align}
\mathbf{a}_{{\rm pr}, i} & =  \mathbf{\nabla} P_i / \rho_i \label{a:pr} \;,\\ 
\mathbf{a}_{{\rm pot}, i} & = \mathbf{(\nabla \times B_i) \times B_i}/ 4 \pi \rho_i \;, \label{a:pot}\\
\mathbf{a}_{{\rm m},i} &= \mathbf{\nabla} \Phi_i ;, \label{a:m}
\end{align}
with the mass density $\rho_i = \mu\,n_i$. Similar to the velocity discussed in Sect.~\ref{subsec:velocities} we determine the mean acceleration as function of number density $n$ by averaging over all cells in the range $\log n-\delta(\log n)/2$ to  $\log n+\delta(\log n)/2$. 

\begin{figure*}
  \centering
    \includegraphics{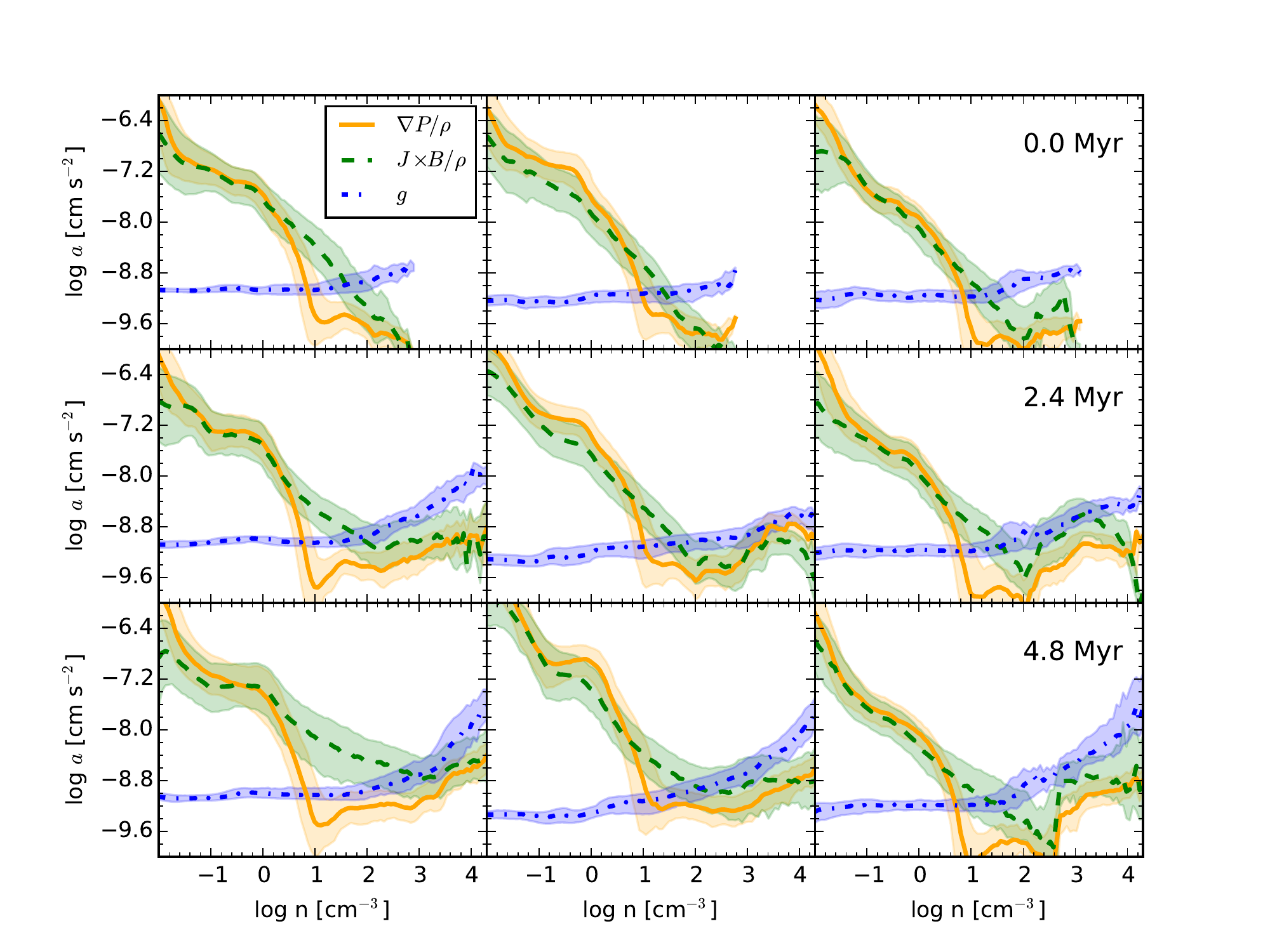}
  \caption{\label{fig:forces}
 Mean accelerations from gravitational potential gradients ({\em blue dash-dot}),
 magnetic forces ({\em
 green dash}), and pressure gradients ({\em orange solid})
 as a function of density in the gas within a ($100$~pc)$^{3}$
 box, centered on the cloud's center of mass for clouds M3 ({\em left}), M4 ({\em center}), and M8 ({\em right}) for three snapshots at $t=0$, 2.5, and
 4.\rf{8}~Myr since the time self-gravity has been turned
 on. Lines show median values in each density bin, while shading shows the central 50\% of values in the bins.} 
\end{figure*}

The result is presented in Figure~\ref{fig:forces}, which shows that all three clouds evolve in qualitatively similar manner. At the lowest densities, magnetic and pressure forces roughly balance each other, while local gravitational forces are unimportant. At intermediate densities, where the gas has radiatively cooled, removing pressure support, magnetic forces dominate. At early times, the highest density regions have been assembled by turbulent flows. Gravitational forces already dominate the accelerations there, once self-gravity is turned on (by our choice of the region) driving the accretion and collapse of the cloud. \rf{The lack of hydrostatic equilibrium in cloud cores was also shown by \citet{girichidis2021} using lower-resolution models of more clouds.} As time passes, the highest density regions become increasingly gravitationally dominated as the cloud continues to accrete and collapse. Thus, these results remain consistent with the picture of magnetically dominated envelopes and gravitationally dominated cores.

\section{Discussion}
\label{sec:observ}

In this discussion we turn to quantities  {that we derive from the three zoom-in clouds}
that can more directly be compared to observations, even if they are not as easily interpretable physically. In Sect.~\ref{subsec:B-n} we focus on the magnetic field strength as a function of density for the diffuse and dense gas in the ISM, with and without gas self-gravity, and compare our measurements with Zeeman observations of magnetic field strengths of MCs in the Galaxy. In Sect.~\ref{subsec:HRO} we compute the relative orientation of the magnetic field to two other quantities: the velocity field (Sect.\ \ref{subsubsec:v-B}), which establishes how the field influences the flows of gas in the diffuse and dense ISM, and the density gradient (Sect.\ \ref{subsubsec:gradn-B}), which quantifies the participation of the field in the structure and collapse of dense MCs. We also address some of the caveats and shortcomings of the current approach. 

\subsection{Density-Magnetic Field Relation}
\label{subsec:B-n}

\begin{figure*}
  \centering
\includegraphics[width=\textwidth]{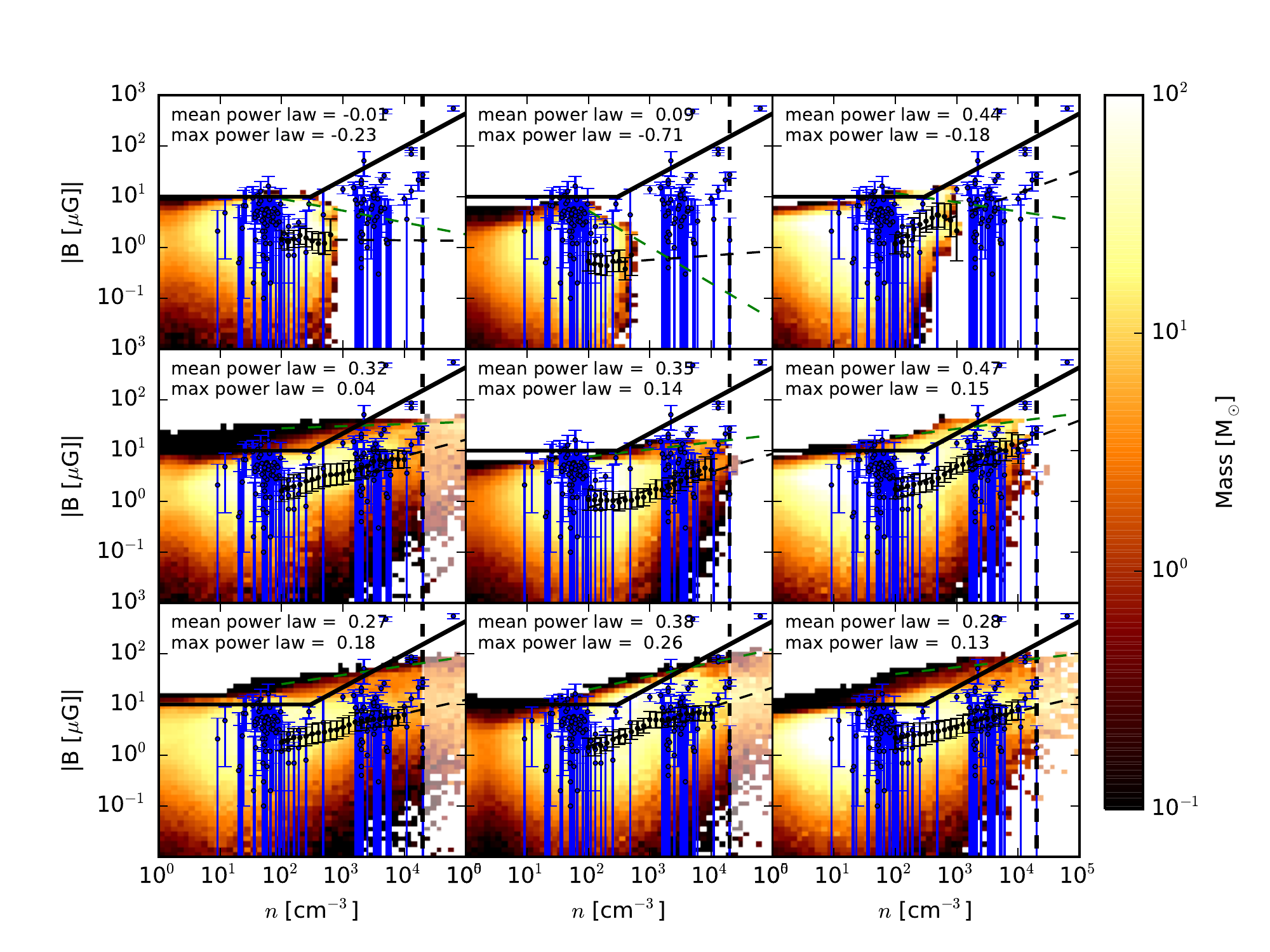}
\caption{Line-of-sight magnetic field strength along the $x$-axis parallel to an arbitrary line of sight in the galactic midplane
versus the gas density in a  $(100$~pc$)^{3}$ volume centered in the cloud's center of mass for
 clouds M3 {\em (left)}, M4 {\em (center)}, and M8 {\em (right)}.
The panels show evolutionary times of 0, 2.9, and 4.\rf{8}~Myr since the
time self-gravity was included from {\em top} to {\em bottom}.
A vertical dashed line marks the maximum resolution at which we resolve the Jeans length with at least four cells.
Blue points show Zeeman observations of line-of-sight magnetic field
strengths with their errors \citep[see references in][]{Crutcher2010MAGNETICANALYSIS}.
The solid black line shows the most probable model of the maximum field strength along an observed line of sight, as a function of density \citep{Crutcher2010MAGNETICANALYSIS}. 
The black dashed line with error bars gives the mean and dispersion of the
simulated values; a linear fit gives the power-law index listed to the
upper left. The green dashed line fits the maximum magnetic field at
each density between the transition point to power-law behavior ($n =
10^2$~cm$^{-3}$) and the Jeans resolution limit ($n = 10^4$~cm$^{-3}$), which also yields
a power-law index.
\label{fig:Bn_relation_M3e3}}
\end{figure*}

\begin{figure*}
    \centering
\includegraphics[width=\textwidth]{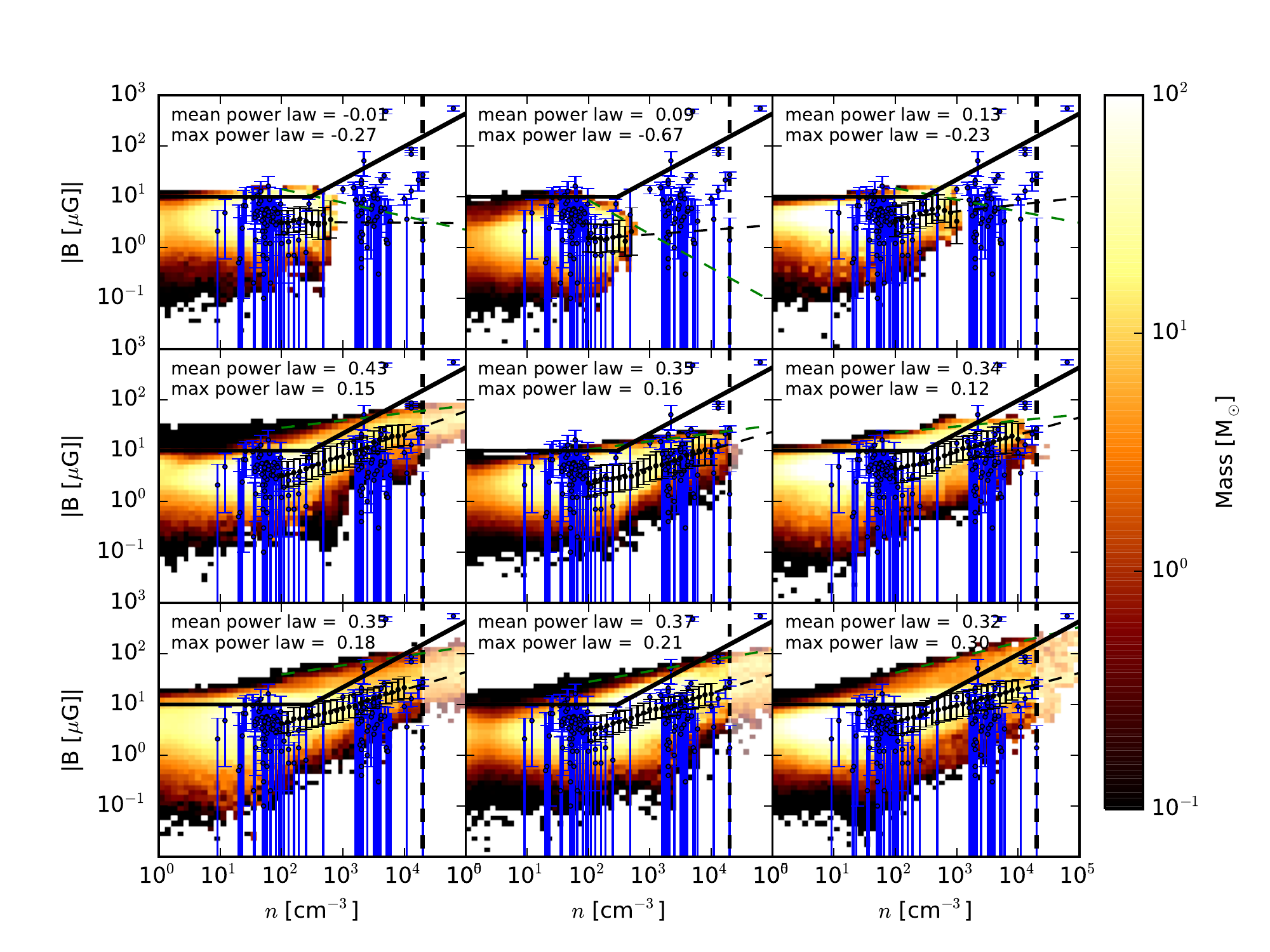}
\caption{Same as Figure~\ref{fig:Bn_relation_M3e3}, but for total
 rather than line-of-sight magnetic field. \label{fig:Btn-relation}}
\end{figure*}

We investigate the correlation between the magnetic field strength and the density in our high-resolution zoom-in models of collapsing clouds using a $\mathbf{B}$-$n$ scatter plot of our model over time. Figures~\ref{fig:Bn_relation_M3e3} and~\ref{fig:Btn-relation} show mass weighted plots of gas density for all cells within the zoom region versus the magnetic field strength, either $x$-component
  (parallel to the line of sight)
or total, measured along an arbitrary line of sight lying in the galactic midplane. These figures also include Zeeman observations towards diffuse and dense clouds compiled by \citet{Crutcher2010Self-consistentObservations} as well as the upper envelope of the magnetic field strength inferred from the observations from that paper.

At $t = 0$, we show the distribution of magnetic field strength as a function of density in the diffuse ISM, as 
described in \citetalias{ibanez-mejia2016}. The bulk of the gas has field strength $|\mathbf{B}_x|$ systematically below $10 ~\mu$G, the upper limit for the magnetic field strength derived from H~{\sc i} observations in the diffuse ISM \citep{Crutcher2012MagneticClouds}. For number densities  $n$ in the range from 1--300~cm$^{-3}$, the magnetic field does not correlate with the gas density, in agreement with the observations. The data show a change in behavior at $n \approx 300$~cm$^{-3}$, where the derived maximum magnetic field strength begins to climb as $|\mathbf{B}|\propto n^{0.65 \pm 0.05}$ \citep{Crutcher2010MAGNETICANALYSIS}.

Indeed, we find a similar increase of the field strength with density. To further quantify this behavior, we fit both
the average and maximum field strength for densities from 10$^2$--$2\times10^{4}$~cm$^{-3}$ using a power law of the form
\begin{equation}
B_{los} = B_{0} \left( \frac{n}{n_{0}}\right)^{\alpha}\;. 
\end{equation} 
As tabulated in Figures~\ref{fig:Bn_relation_M3e3} and \ref{fig:Btn-relation}, we find  exponents $\alpha < 0.4$, which are shallower than expected theoretically for either diffusion-dominated or freely-collapsing clouds, and which are also shallower than what is inferred from the observations by \citet{Crutcher2010MAGNETICANALYSIS}. 

This behavior could result from two effects. First, as we approach the Jeans resolution limit, numerical diffusion becomes important in limiting field growth. Higher resolution models refined with more cells per Jeans length \citep{sur2010} may yield higher field strengths at high densities. Second, our models do not include any treatment of radiative transfer or chemical abundance variations that could selectively emphasize particular regimes. Indeed the shallow slopes {of the maximum field (fits to the green dashed lines in Figs.~\ref{fig:Bn_relation_M3e3} and~\ref{fig:Btn-relation})} result in part from {maximum field values that are higher than predicted by the \citet{Crutcher2010MAGNETICANALYSIS} analysis} at intermediate densities around $n \sim 10^3$~cm$^{-3}$. However, the observations are quite sparse in this regime due to a lack of chemical tracers, so the behavior {in this regime} is {actually quite} weakly constrained.

Models of magnetized cloud collapse suggest that the turbulence during collapse does drive a small-scale dynamo that produces field growth \cite[][]{subramanian2005, schleicher2010, federrath2011b, schober2012b}.
However, \citet{xu2020} argue that at saturation of the dynamo, reconnection diffusion reduces further growth rates below the ideal MHD limit, {possibly} consistent with our simulation results. Further observational investigation of the intermediate density region {with $n \simeq 10^3$~cm$^{-3}$} might demonstrate this behavior.

\subsection{Histogram of Relative Orientations}
\label{subsec:HRO}

We examine the morphology of the magnetic field with respect to the cloud's density structure, velocity field and gravitational forces using the histogram of relative orientations (HRO) as introduced by \citet{soler2013} to studies of the ISM \cite[see also][]{soler2019, soler2020}.  The HRO is a diagnostic derived from the machine vision technique of histograms of oriented gradients \citep{freeman1994,dalal2005}. The original technique examined the distribution of orientations of gradients in two-dimensional images, while the HRO compares the orientation of a gradient to that of a vector field, possibly in three dimensions.

We extend the standard analysis of the HRO method by working in three dimensions (position-position-position space), and by not only studying the relative orientation of the density gradient to the magnetic field, but also the orientation of the density gradient with respect to the velocity field.  The density structure of the cloud can be characterized by the density gradient. Use of the gradient allows local comparison of the orientation of the magnetic field to iso-density contours. The HRO is computed by measuring the relative angle between the local density gradient $\mathbf{\nabla} n$ and the magnetic field $\mathbf{B}$ as
\begin{eqnarray}
	\phi_{\hat{n}\hat{B}} = 
 \tan^{-1}\left( 
 \frac{|\mathbf{\nabla} n \mathbf{\times} \mathbf{B}| }{\mathbf{\nabla} n \mathbf{\cdot} \mathbf{B}} 
  \right)\;.
	\label{eq:hro}
\end{eqnarray}
We can also measure the relative angle between the magnetic field and the velocity $\phi_{\hat{v}\hat{B}}$ using the velocity vector in place of the density gradient in Eq.~(\ref{eq:hro}). Together with the total energy measurements and the characteristic velocities, this diagnostic can reveal whether gas flows are restricted to follow field lines, or whether magnetic fields are just being advected by the turbulence. Values of $\phi_{\hat{v}\hat{B}}$ consistently close to zero suggest that the velocities are constrained by the field, while a broad range of values suggests that field is being advected by the turbulence kinematically.  The latter would also be likely if the kinetic energy exceeds the gravitational energy, and the rms velocity exceeds the Alfv\'en velocity. However then the scale of the turbulence determines the scale of magnetic field coherence.

Once the relative angles have been computed everywhere in the model, we collect this data into a histogram to determine if there are any preferred orientation angles. We proceed to bin the results with respect to the density, in order to explore the behavior of these relative orientations for a wide range of environments. Given that we perform our study of the HRO for physical scales over the range 0.1--100~pc in an AMR simulation, we do not force our density bins to have equal numbers of pixels, but rather force a minimum number of pixels per bin ($> 10^3$)  to produce a well-sampled distribution of angles.

%
\begin{figure*}[t]
\centering 
\begin{interactive}{animation}{M3_HRO.mp4}
\includegraphics[width=0.7\textwidth]{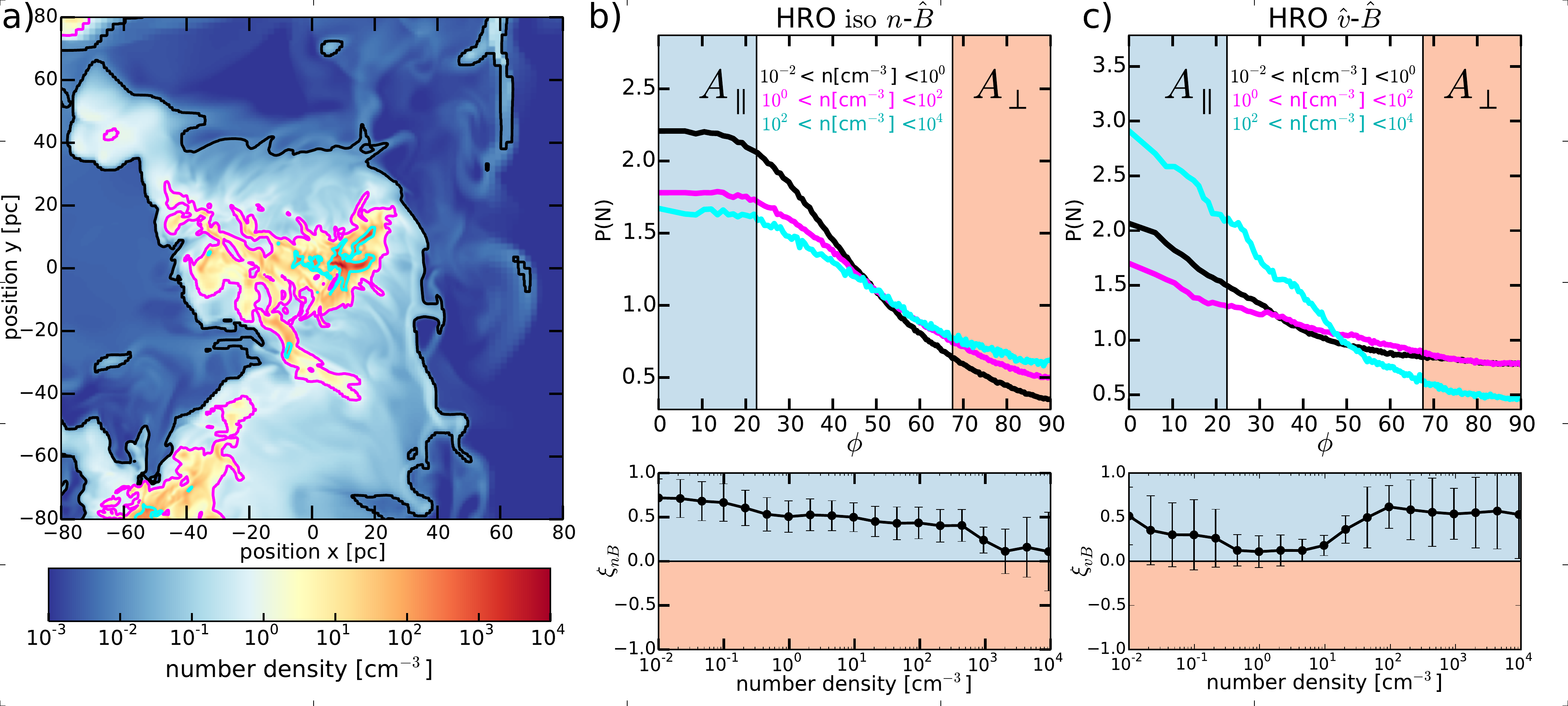}
\end{interactive}
\caption{ Animations over the first 5~Myr after gravitational collapse
  for cloud M3 
  of (a) density slices with contours at $10^{-2}, 1$~and $100$~cm$^{-3}$, and probability density functions of the relative orientations
 between the magnetic field $\mathbf{\hat{B}}$ and (b) the local density gradient, $\mathbf{\nabla} n$ and (c)
 the velocity $\mathbf{v}$, as a function of the given gas
 density ranges, within a ($100$~pc)$^{3}$ box centered on \rf{the}
 cloud's center of mass. The histogram shape parameter $\xi$
 given by Equation~(\ref{eq:xi}) with error bars from
 Equation~(\ref{eq:sig_xi}) is shown as
 a function of density within narrower bins for \rf{the}
 distribution. The regions contributing to $\xi$ are colored
 for clarity: parallel dominance in blue and perpendicular dominance
 in orange. The still figure published is at the final time for the
 each model.  \rf{The animation shows the evolution of the field in the highest density gas from parallel to neutral or perpendicular orientation to the density gradient as collapse proceeds.}
\label{fig:Angles_nB1}} 
\end{figure*}
\begin{figure*}
\begin{interactive}{animation}{M4_HRO.mp4}
\includegraphics[width=0.7\textwidth]{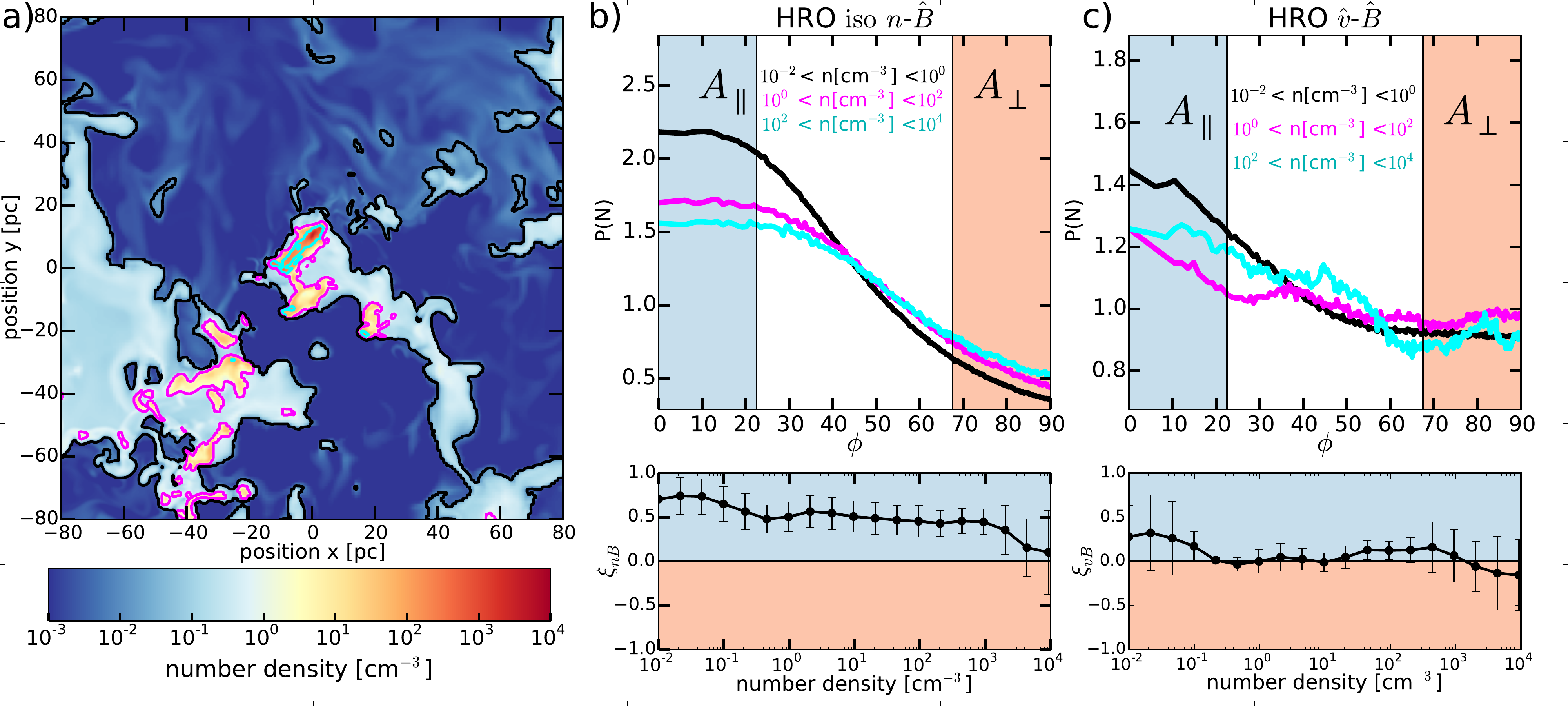}
\end{interactive}
\caption{\rf{Same as Figure~\ref{fig:Angles_nB1} for model M4, showing similar behavior.} \label{fig:Angles_nB2}}
\end{figure*}
\begin{figure*}
\begin{interactive}{animation}{M8_HRO.mp4}
\includegraphics[width=0.7\textwidth]{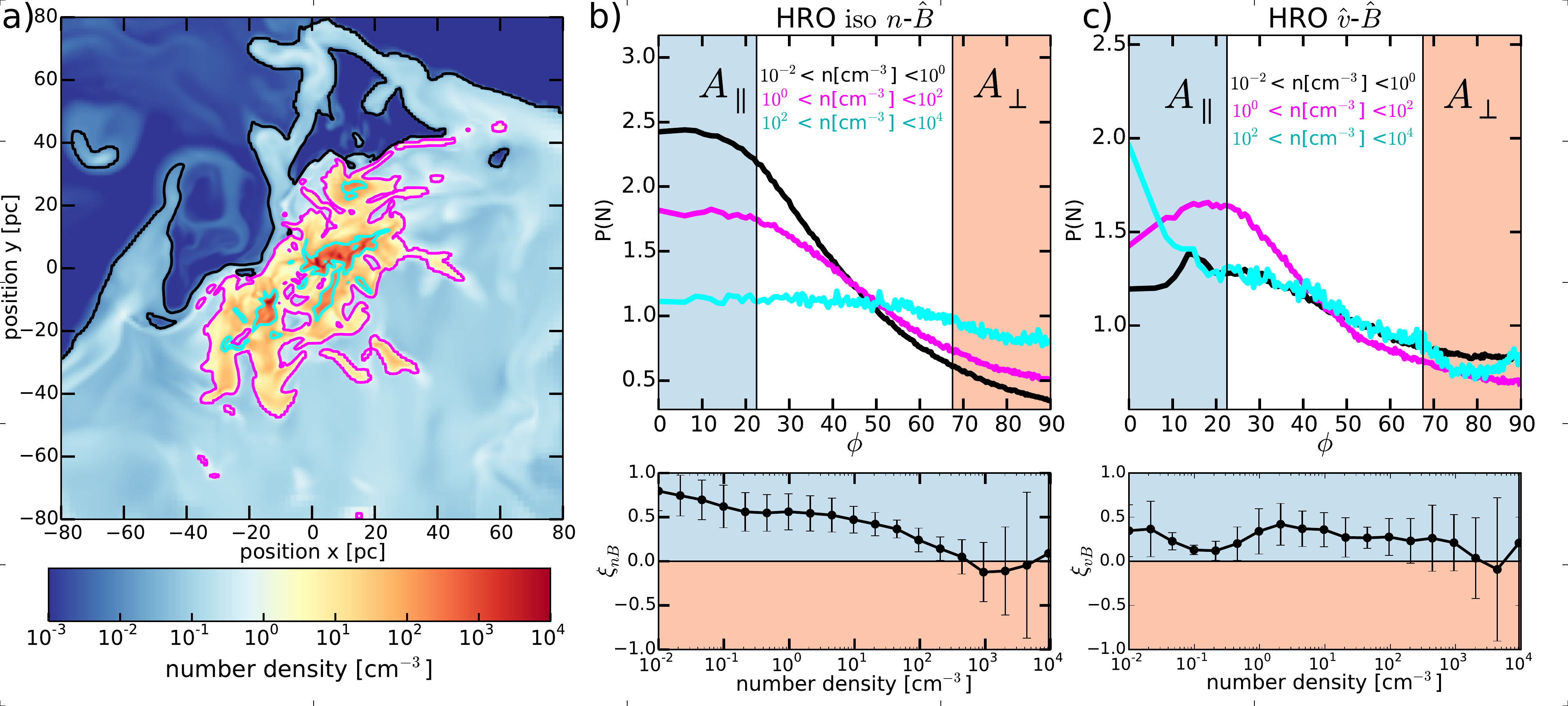} 
\end{interactive} 
\caption{\rf{Same as Figure~\ref{fig:Angles_nB1} for model M8, showing similar behavior.} \label{fig:Angles_nB3}}
\end{figure*}

To characterize the shape of the HRO in each bin we employ the HRO shape parameter also used by \citet{seifried2020}
\begin{equation} \label{eq:xi}
 \xi = \left(A_{||} - A_{\perp}\right) / \left(A_{||} + A_{\perp}\right)\;,
\end{equation}
where $A_{||}$ is the area under the histogram for $1 > \cos \phi > 0.75$ and $A_{\perp}$ is the area under the histogram for $0.25 > \cos \phi > 0$. The relevant ranges are indicated in Figure\rf{s}~\ref{fig:Angles_nB1}--\ref{fig:Angles_nB3} with blue and orange shading. A positive value of $\xi$ corresponds to parallel dominance, while a negative value corresponds to a largely perpendicular orientation. We compute the variance in $\xi$ within each density bin by taking the standard deviation $\sigma_{||}$ of the HRO values across the range of angles considered for $A_{||}$, and a corresponding value for $\sigma_{\perp}$. Then
\begin{equation} \label{eq:sig_xi}
 \sigma_{\xi} = \left[4 \frac{A_{\perp}^2 \sigma_{||}^2 + A_{||}^2 \sigma_{\perp}^2}{\left(A_{||} + A_{\perp}\right)^4}\right]^{1/2}\,.
\end{equation}
Finally, as we are interested in the dynamical evolution of the cloud as it collapses, we compute the time evolution of the HRO.

\subsubsection{Density Gradient-Magnetic Field Relative Orientation}
\label{subsubsec:gradn-B}

We next use this characterization to describe the relative orientation of the magnetic field with respect to the density gradients in and around the cloud. \rf{Panels (b) of} Figure\rf{s}~\ref{fig:Angles_nB1}--\ref{fig:Angles_nB3} show the evolution over time of the HRO as a function of gas density. For  $n > 100$~cm$^{-3}$ the flow becomes super-Alfv\'enic {as collapse proceeds}. The relative orientation {at these higher densities} shows a transition from predominantly parallel {with $\zeta_{nB} > 0.5$} to much more random or even perpendicular {with $\zeta_{nB} \lesssim 0$ and variance $\sigma_{\zeta} > 1$ as the field gets swept along by the flow.}

This result agrees with previous applications of the HRO to simulated clouds by \citet{soler2013}, \citet{soler2017}, \citet{seifried2020}, \rf{and \citet{girichidis2021}} using different approaches to cloud formation or structure. The general observation that fields typically shift from parallel to perpendicular during the collapse of dense structures has a long history \citep{heitsch2001, ostriker2001, zli2004,nakamura2008, collins2011, hennebelle2013, soler2013, chen2015a, li2015, chen2016, zamora-aviles2017, mocz2018, chen2020, seifried2020, girichidis2021}. The critical density for the transition furthermore agrees with the density at which the field starts to increase with density (Figures~\ref{fig:Bn_relation_M3e3} and \ref{fig:Btn-relation}), as would be expected from a collapse-dominated flow. 

\subsubsection{Velocity-Magnetic Field Relative Orientation}
\label{subsubsec:v-B}
In order to explain the flat behavior of the magnetic field-density relation for densities below $n<100$~cm$^{-3}$, it has been suggested that gas flows preferentially along field lines. 
%
\begin{figure*}[t]
\centering 
\includegraphics[width=0.49\textwidth]{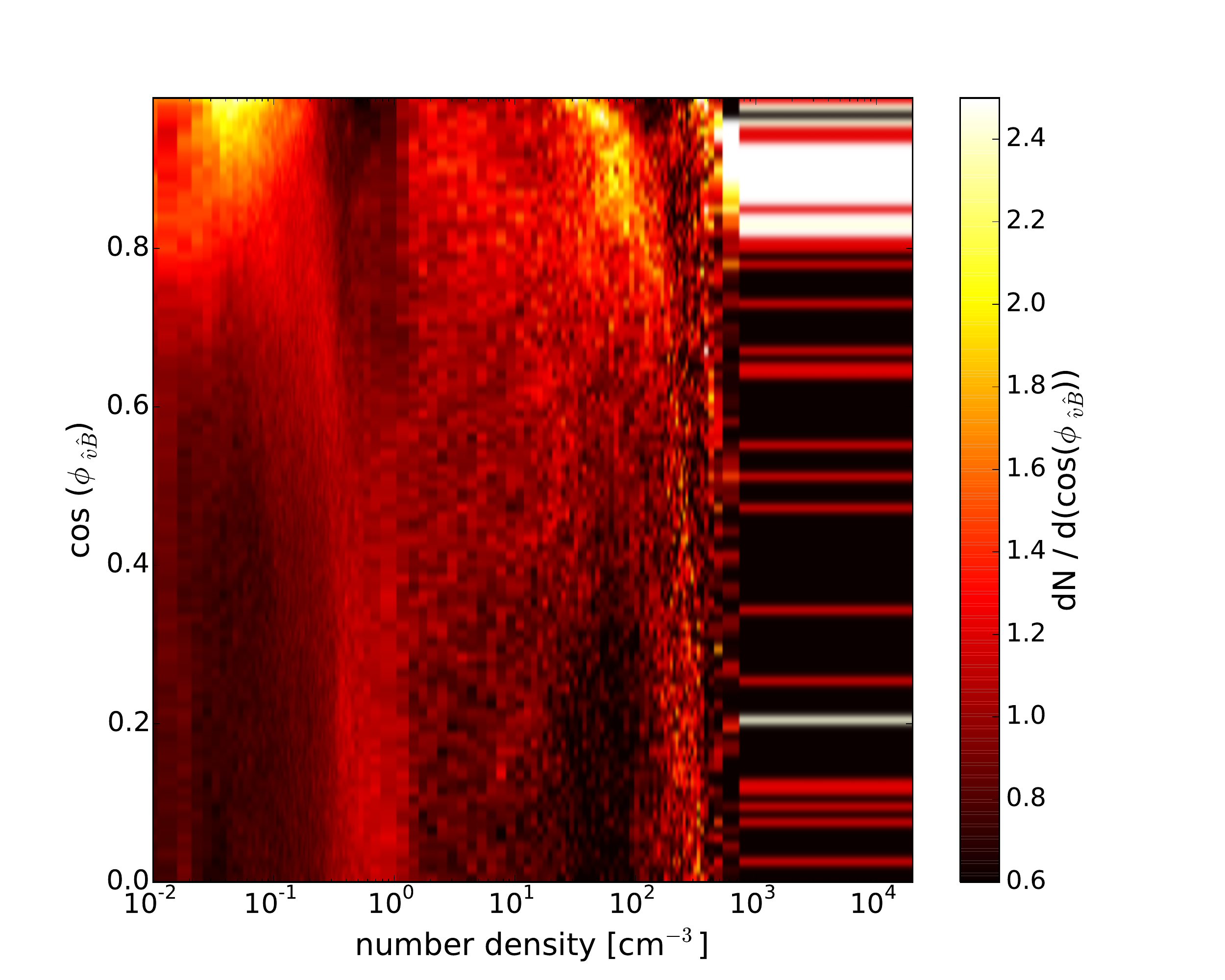} 
\includegraphics[width=0.49\textwidth]{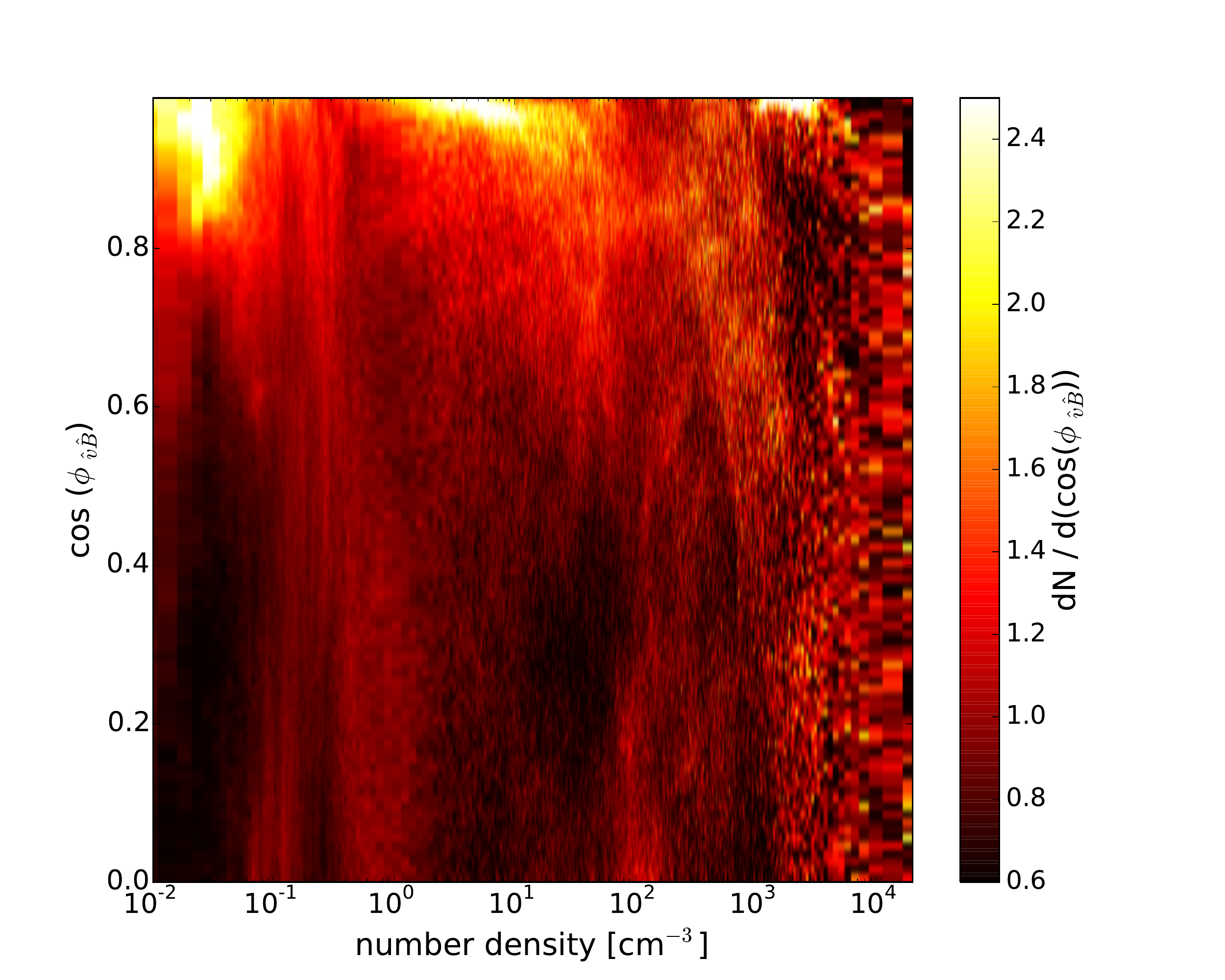} \\
\includegraphics[width=0.49\textwidth]{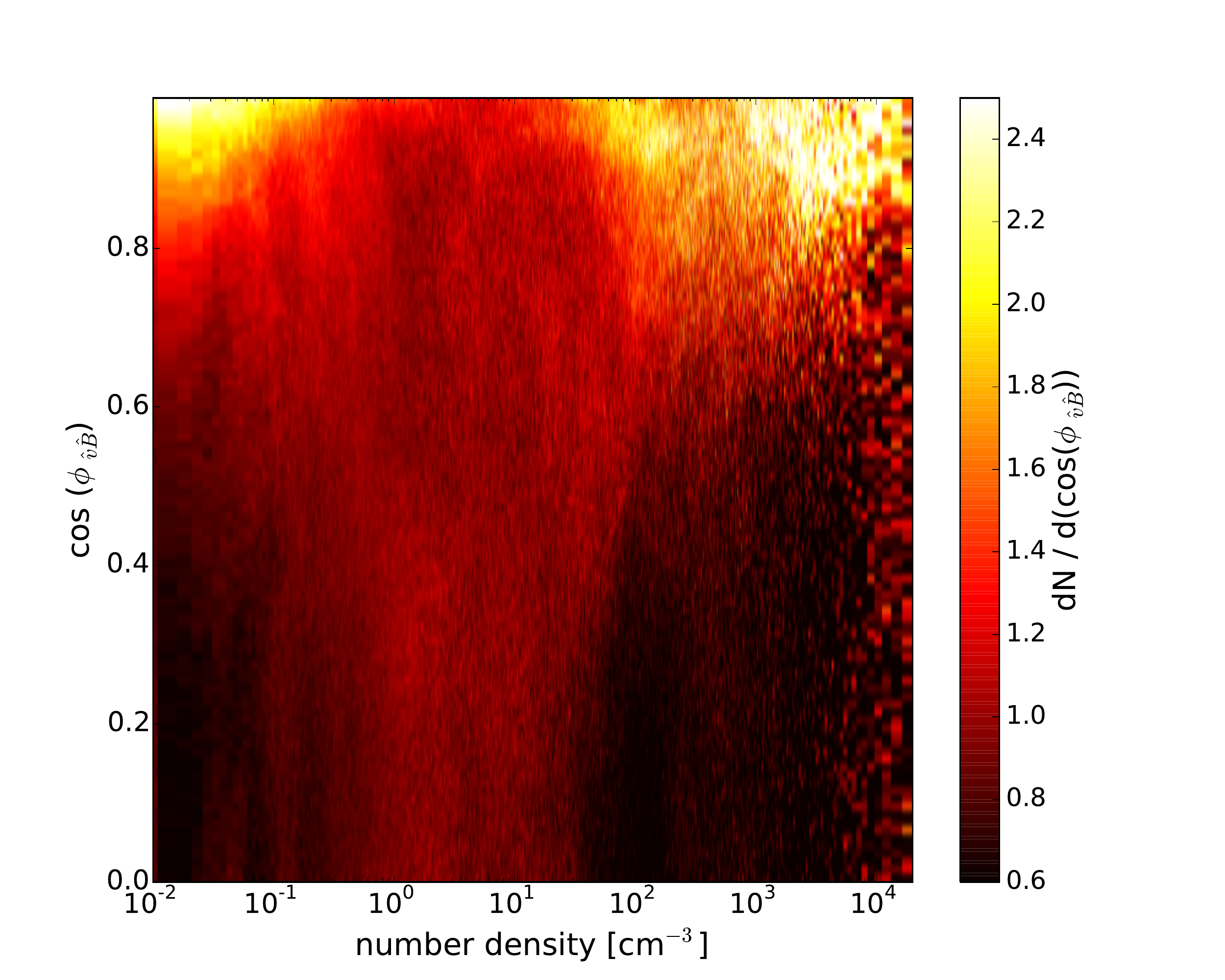} 
\includegraphics[width=0.49\textwidth]{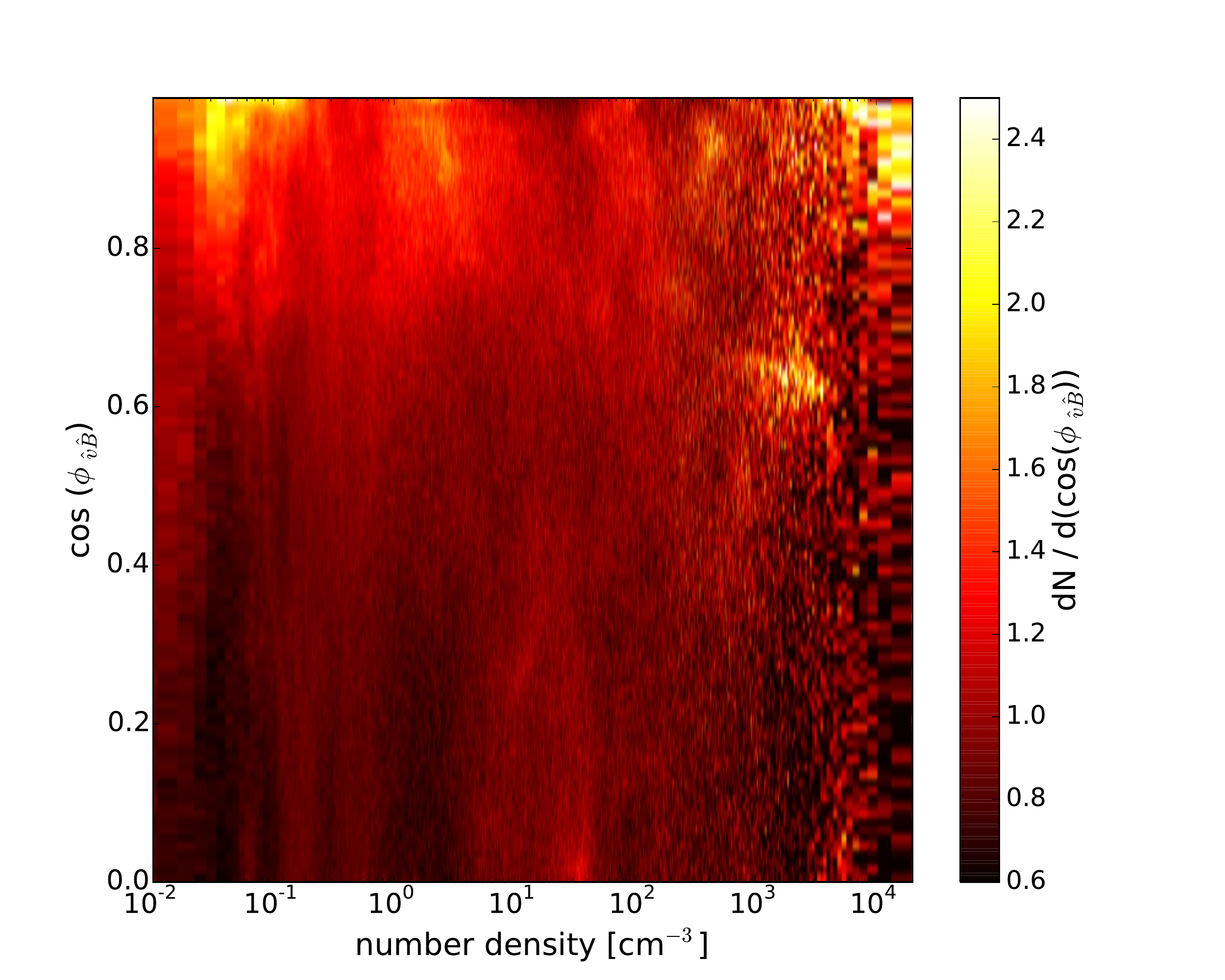} 
\caption{Probability density function of the relative orientations
 between the local velocity, $\hat{v}$, and the magnetic field,
 $\hat{B}$, as a function of gas density, within a ($100$~pc)$^{3}$
 box centered on the center of mass of cloud M3.
The panels correspond to evolutionary times of 0, 2, 4, and 6~Myr
since the moment self-gravity was included. (At the earliest time
there is no high density gas, so the final bin covers a wide range of
densities.) \label{fig:Angles_vB}} 
\end{figure*}
\rf{Panels (c) of} Figure\rf{s}~\ref{fig:Angles_nB1}--\ref{fig:Angles_nB3} show variation of the HRO shape parameter for the angle between field and velocity $\xi_{\hat{v}\hat{B}}$ over time, while Figure~\ref{fig:Angles_vB} shows the actual distribution of the cosine of the angle between the direction of the velocity $\hat{\mathbf{v}}$, and the magnetic field, $\hat{\mathbf{B}}$, as a function of density, at four different evolutionary stages of cloud M3. At all stages of evolution the magnetic field indeed tends to be oriented parallel to the flow velocity,  $\cos \phi_{\hat{v}\hat{B}} = ${1}, for densities below 100~cm$^{-3}$. 

However configurations with the field perpendicular to the flow still constitute a substantial fraction of the volume, as even the lowest probability densities exceed 0.6, a quarter of the maximum value. The probability density of relative orientations is almost flat for number densities $0.5 < n < 50$. This density range corresponds to transition gas in a thermally unstable phase of the ISM. Gas remains in this density range for only a short time, as it consists mostly of shocked gas, rapidly cooling towards the cold, dense phase of the ISM.

Large fluctuations in the relative distribution of the orientations are observed at $t \sim 4$~Myr. This is likely caused by the explosion of a SN within the analyzed volume at $t=2.62$~Myr, whose blast wave hits the cloud after a transit time of $\sim 1$~Myr. The SN remnant expands into an inhomogeneous density distribution, producing turbulent motions both parallel and perpendicular to the magnetic field, resulting in the predominantly flat behavior of the probability seen.

The alignment between the velocity and the magnetic field might not only be caused by strong magnetic fields. As discussed by \citet{padoan1999} there are two opposite processes that can cause alignment: dynamical alignment, occurring when the field is strong enough to restrict the gas flows along field lines, and kinematic alignment, occurring when the field is swept up by the gas flows, forcing alignment. It is likely that for densities $n< 100$~cm$^{-3}$, the velocity and magnetic field are dynamically aligned, as the flows are trans-Alfv\'enic (see Sect.~\ref{subsec:velocities}), at least when the system has not been recently perturbed by a nearby SN explosion. In this regime, the magnetic tension can restrict gas flows perpendicular to the magnetic field.

For densities $n>100$~cm$^{-3}$, the angle $\phi_{\hat{v}\hat{B}}$ shows large fluctuations as a function of time and density. This regime is less likely to be affected by random turbulence in the environment, but on the other hand is most likely to be affected by hierarchical
gravitational contraction, as gravity is the dominant form of energy here (see Sect.~\ref{subsec:Energetics}). As the cloud contracts, it can compress the magnetic field. At $t=2$~Myr, $\phi_{\hat{v}\hat{B}}$ transitions at these densities from a preferentially aligned flow into a random distribution of the alignment. At $t=4$~Myr, $\phi_{\hat{v}\hat{B}}$ shows some alignment between the velocity and the magnetic field, which is most certainly caused by kinematic alignment of the collapsing gas, as the flow here is super-Alfv\'enic. Later, at $t=6$~Myr, the relative distribution of the angles is very random, with more of the flow aligned with the field, but with large fractions of the gas having oblique relative orientations. This erratic behavior of the relative angles shows that when gravity controls the dynamics of the gas, the magnetic field is carried along with the contraction.

\subsection{Caveats}
\label{subsec:caveats}

Our results apply in detail only to the earliest evolution of dense
clouds, as we have no star formation or feedback models in our
simulations. \rf{As a result, we do not follow the expected prompt cloud destruction from feedback that must occur for gravitationally collapsing clouds not to exceed the observed star formation rate, as we discussed in Sect.~\ref{sec:introduction}.}

Even during \rf{the period of collapse}, we are not following the
detailed chemistry and ionization structure of the clouds but rather
relying on a tabulated temperature-dependent cooling prescription to
follow the thermodynamics of cloud formation. For comparison,
\citet{seifried2017} \rf{and \citet{girichidis2021}} included chemistry, but not early stellar
feedback, while \citet{grudic2021} included both, reaching overall
similar results on collapse and field structure. Our approach does
allow us to focus in detail on the force structure and energetics of
the collapsing cloud.

Our most significant numerical limitation is that we only resolve Jeans lengths with a minimum resolution of four cells, as expressed by the \citet{truelove1997} criterion \cite[see also][]{heitsch2001}. This may be inadequate to resolve the turbulence and field evolution within the collapsing regions of the clouds \citep[e.g.][]{sur2010, federrath2011}, which could limit the applicability of our conclusions about the dynamics of the densest regions of the clouds.  \rf{It is worth noting that \citet{girichidis2021} did a resolution study from 2~pc to 0.25~pc grid resolutions confirming convergence of the major results on field alignment and gravitational collapse.}

The study of field angles has all been done for three-dimensional structures, where the dynamics remains unobscured by projection effects. These have been shown by \citet{soler2019}, \citet{seifried2020}, \rf{and \citet{girichidis2021}} using detailed radiative transfer models to significantly impact the ability to deduce information about field structure from observations.

\section{Summary and Conclusion}
\label{sec:conclusions}

We have presented here three-dimensional MHD simulations of dense clouds formed in a supernova-driven, stratified, turbulent, galactic environment, collapsing under the action of their own self-gravity. These calculations were originally presented in \citetalias{ibanez-mejia2016} and \citetalias{ibanez-mejia2017}. We find that:
\begin{itemize}
\item Dense clouds with $n > 100$~cm$^{-3}$ have super-Alfv\'enic rms  velocity dispersions, caused by fast turbulent motions driven by hierarchical gravitational contraction, suggesting that the fields are not dynamically important in this regime. In contrast, diffuse gas with $n < 100$~cm$^{-3}$ in the envelopes of MCs and the diffuse ISM exhibits trans-Alfv\'enic velocity dispersion \citep{boulares1990}. This suggests the medium is magnetically supported against gravitational collapse \citep{Elmegreen2007OnClouds}, and its flow is generally constrained to follow field lines \citep{padoan1999,heiles2005}, although substantial deviations to this simple picture are seen in our simulations.
\item Examination of the relative energy in the clouds shows that kinetic, magnetic and thermal energy are in equipartition in the diffuse medium with $n < 1$~cm$^{-3}$, kinetic energy dominates in the range of roughly $1 < n < 100$~cm$^{-3}$, whereas gravitational energy dominates at densities $n > 100$~cm$^{-3}$. This is consistent with collapse dominated clouds.
 \item Direct analysis of the acceleration terms contributing to the momentum equation emphasizes that thermal pressure gradients and Lorentz forces are equally important in the diffuse medium, magnetic fields dominate at intermediate densities, but self-gravity determines the flow properties at high densities.
 \item Random turbulent motions in the diffuse ISM and gravitational contraction of dense structures maintain the magnetic field strength at magnitudes consistent with those reported in observations \citep{Crutcher2012MagneticClouds}. The behavior of the magnetic field strength in relation to density also appears generally consistent, with no correlation between the field strength and gas density for densities $n<100$~cm$^{-3}$, and a positive correlation above that density. However, the average $\mathbf{B}$-$n$ relation in our model has a significantly shallower slope than the observed relation between the maximum line-of-sight field strength and the derived density.
\item Three-dimensional HROs of magnetic field vs.\ density gradient in our models indicate the transition from parallel fields in the diffuse medium and unbound cloud envelopes to random orientations---however, with some tendendy towards perpendicular configurations---in bound high density cloud cores. 
\item While gas flows are {\em{dynamically aligned}} along the magnetic field at low densities, oblique supersonic shocks produce strong density fluctuations that lead to significant volumes with the field more perpendicular to the density gradient than expected for
  pure dynamical alignment.
  
\end{itemize}

\begin{acknowledgments}
 We thank J. Soler, L. Fissel, and B. Burkhart for stimulating conversations\rf{ and the anonymous referee for a detailed and useful report}.
M-MML received support from US NSF grant AST18-15461, and thanks the A. von Humboldt-Stiftung for travel support. JCI-M was partly supported by the Deutsche Forschungsgemeinschaft (DFG) via the Collaborative Research  Center SFB 956 {\em Conditions and Impact of Star Formation} (subproject C5) and the DFG Priority Program 1573 {\em The physics of the interstellar medium}.  RSK acknowledges support from the DFG via the Collaborative Research Center (SFB 881, Project-ID 138713538) {\em The Milky Way System} (sub-projects A1, B1, B2 and B8); from the Heidelberg Cluster of Excellence (EXC 2181 - 390900948) {\em STRUCTURES: A unifying approach to emergent phenomena in the physical world, mathematics, and complex data}, funded by the German Excellence Strategy; and from the European Research Council in the ERC Synergy Grant {\em ECOGAL -- Understanding our Galactic ecosystem: From the disk of the Milky Way to the formation sites of stars and planets} (project ID 855130). The project in part made use of computing resources provided by the state of Baden-W\"{u}rttemberg and the German Research Foundation (DFG) through grants INST 35/1134-1 FUGG and  INST 35/1314-1 FUGG. 
\end{acknowledgments}

\software{yt \citep{turk2011}, Flash \citep{fryxell2000}, numpy \citep{harris2020},
 matplotlib \citep{hunter2007}}

\bibliographystyle{aasjournal}
\bibliography{Mendeley,maclow,klessen}


\end{document}